\documentclass[10pt,twocolumn,letterpaper]{article}

\usepackage[final]{cvpr}      

\usepackage{amsmath}
\usepackage{amssymb}
\usepackage{booktabs}

\usepackage{times,graphicx,amsmath,amssymb,booktabs,tabulary,multirow,overpic,xcolor}
\usepackage{colortbl}
\usepackage{makecell}
\usepackage{float}

\definecolor{sh_gray}{rgb}{0.84,0.84,0.84}
\definecolor{sh_gray2}{rgb}{1,0.89,0.75}
\definecolor{color3}{rgb}{0.95,0.95,0.95}
\definecolor{color4}{rgb}{0.96,0.96,0.86}
\definecolor{color5}{rgb}{0.90,0.90,0.90}

\usepackage[pagebackref,breaklinks,colorlinks]{hyperref}

\usepackage[capitalize]{cleveref}
\usepackage{multirow}
\crefname{section}{Sec.}{Secs.}
\Crefname{section}{Section}{Sections}
\Crefname{table}{Table}{Tables}
\crefname{table}{Tab.}{Tabs.}
\usepackage{algorithm}
\usepackage{algorithmic}

\def\cvprPaperID{***} 
\def\confName{CVPR}
\def\confYear{2022}

\begin{document}

\title{Mask-guided Spectral-wise Transformer for \\ Efficient Hyperspectral Image Reconstruction}

\author{%
	Yuanhao Cai $^{1,2,*}$, Jin Lin $^{1,2,}$\thanks{The first two authors contribute equally. Code will be made public.} ~, Xiaowan Hu $^{1,2}$, Haoqian Wang $^{1,2}$, \\ Xin Yuan $^3$, Yulun Zhang $^4$, Radu Timofte $^4$, and Luc Van Gool $^4$ \\
		$^{1}$ Shenzhen International Graduate School, Tsinghua University, \\ $^2$  Shenzhen Institute of Future Media Technology, $^3$ Westlake University, $^4$ ETH Z\"{u}rich
}
\maketitle

\begin{abstract}
\vspace{-4mm}
Hyperspectral image (HSI) reconstruction aims to recover the 3D spatial-spectral signal from a 2D measurement in the coded aperture snapshot spectral imaging  (CASSI) system. The HSI  representations are highly similar and correlated across the spectral dimension. Modeling the inter-spectra interactions is beneficial for HSI reconstruction. However, existing CNN-based methods show limitations in capturing spectral-wise similarity and long-range dependencies. Besides, the HSI information is modulated by a coded aperture (physical mask) in CASSI. Nonetheless, current algorithms have not fully explored the guidance effect of the mask for HSI restoration. In this paper, we propose a novel framework, Mask-guided Spectral-wise Transformer (MST), for HSI reconstruction. Specifically, we present a Spectral-wise Multi-head Self-Attention (S-MSA) that treats each spectral feature as a token and calculates self-attention along the spectral dimension. In addition, we customize a Mask-guided Mechanism (MM) that directs S-MSA to pay attention to spatial regions with high-fidelity spectral representations. Extensive experiments show that our MST significantly outperforms state-of-the-art (SOTA) methods on simulation and real HSI datasets while  requiring dramatically cheaper computational and memory costs.
\end{abstract}

\vspace{-8.5mm}
\section{Introduction}
\vspace{-2mm}
\begin{figure}[h]
	\begin{center}
		\begin{tabular}[t]{c} \hspace{-3.8mm} 
			\includegraphics[width=0.48\textwidth]{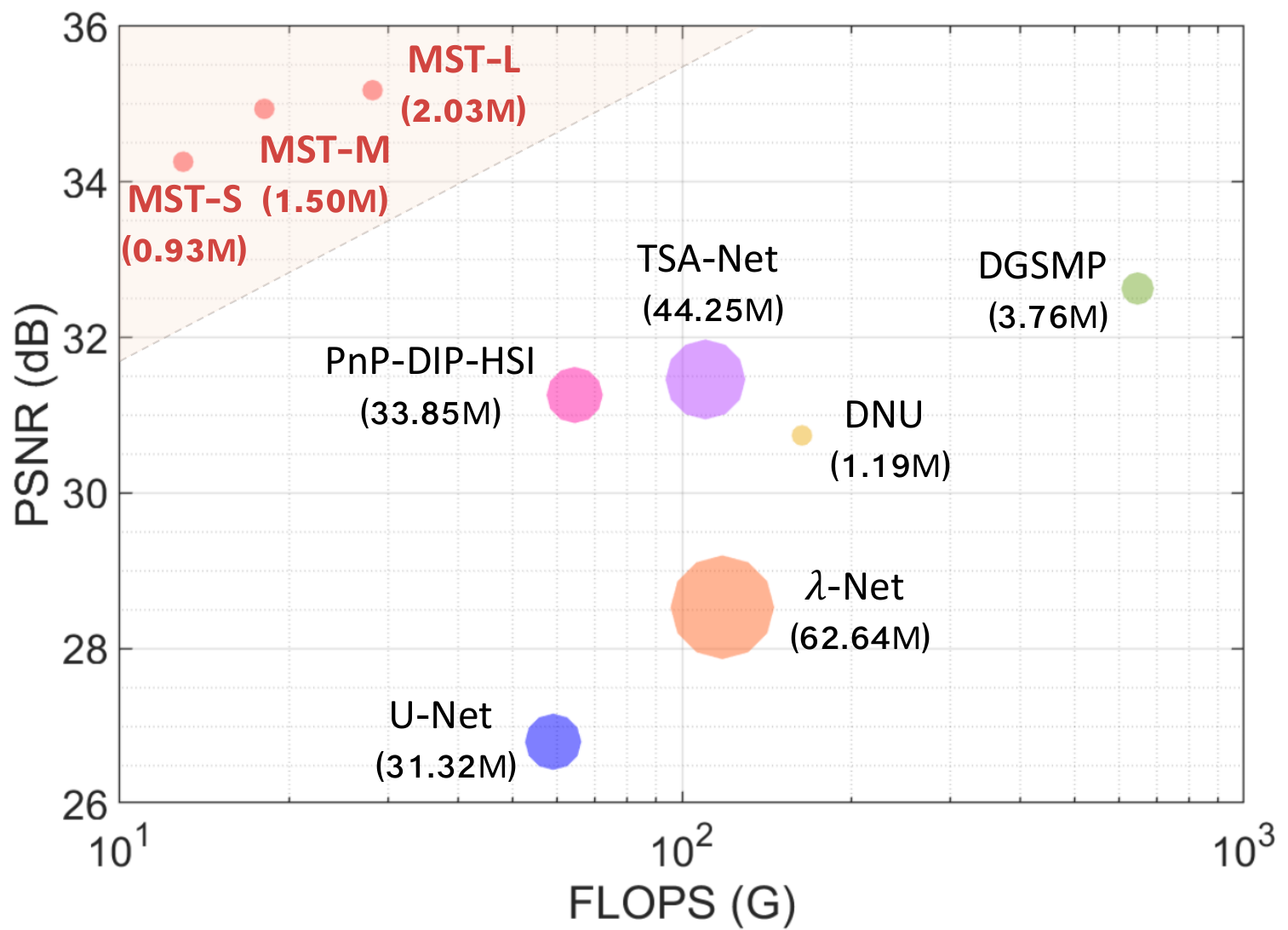}
		\end{tabular}
	\end{center}
	\vspace{-6.5mm}
	\caption{\small PSNR-Params-FLOPS comparisons of CNN-based HSI reconstruction algorithms. The vertical axis is PSNR (in dB performance), the horizontal axis is FLOPS (computational cost), and the circle radius is Params (memory cost). Our proposed Mask-guided Spectral-wise Transformers (MSTs) outperform previous methods while requiring significantly cheaper FLOPS and Params. }
	\label{fig:teaser}
	\vspace{-4.5mm}
\end{figure}
Hyperspectral imaging refers to multi-channel imaging where each channel captures the information at a specific spectral wavelength for a real-world scene. Generally, hyperspectral images (HSIs) have more spectral bands than normal RGB images to store richer information and delineate more detailed characteristics of the imaged scene. Relying on this property, HSIs have been widely applied to many computer vision related tasks, \emph{e.g.}, remote sensing~\cite{rs_1,rs_2,rs_3,rs_4}, object tracking~\cite{ot_1,ot_2,ot_3}, medical image processing~\cite{mi_1,mi_2,mi_3}, \emph{etc.} To collect HSIs, conventional imaging systems with spectrometers scan the scenes along the spatial or spectral dimension, usually requiring a long time. Therefore, these traditional imaging  systems are unsuitable for capturing and measuring dynamic scenes. Recently, researchers have used snapshot compressive imaging (SCI) systems to capture HSIs. These SCI systems compress information of snapshots along the spectral dimension into one single 2D measurement~\cite{Yuan_review}. Among current existing SCI systems~\cite{sci_1,sci_2,sci_3,sci_4,sci_5,sci_6}, the coded aperture snapshot spectral imaging (CASSI)~\cite{tsa_net,sci_2} stands out and forms one  promising mainstream research direction.

Based on CASSI, a large number of reconstruction algorithms have been proposed to recover the 3D HSI cube from the 2D measurement. Conventional model-based methods adopt hand-crafted priors such as sparsity~\cite{sparse_1,sparse_2,sparse_3}, total variation~\cite{tv_1,tv_2,gap_tv}, and non-local similarity~\cite{desci,non_local_1,non_local_2} to regularize the reconstruction procedure. However, these methods need to tweak parameters manually, resulting in poor generalization ability, unsatisfactory reconstruction quality, and slow restoration speed. With the development of deep learning, HSI reconstruction has witnessed a significant progress. Deep convolutional neural network (CNN) applies a powerful model to learn the end-to-end mapping function from the 2D measurement to the 3D HSI cube. Although impressive results have been achieved, CNN-based methods~\cite{lambda,tsa_net,gsm,gapnet,tra_1} show limitations in modeling the inter-spectra similarity and long-range dependencies. Besides, the HSIs are modulated by a physical mask in CASSI. Nonetheless, previous methods~\cite{tsa_net,resu,gapnet} simply conduct an inner product between the mask and the shifted version of the measurement. This scheme corrupts the input HSI representations and does not fully explore the {\em guidance effect of the mask}, leading to limited improvement.

In recent years, the natural language processing (NLP) model, Transformer~\cite{vaswani2017attention}, has been introduced into computer vision and outperformed CNN methods in many tasks. The Multi-head Self-Attention (MSA) module in Transformer excels at capturing non-local similarity and long-range dependencies. This advantage  provides a possibility to address the aforementioned limitations of CNN-based methods in HSI reconstruction. However, directly applying the original Transformer may be unsuitable for HSI restoration due to the following reasons. \textbf{Firstly}, original Transformers learn to capture the long-range dependencies in spatial wise but the representations of HSIs are spectrally highly self-similar. In this case, the inter-spectra similarity and correlations are not well modeled. Meanwhile, the spectral information is spatially sparse. Capturing spatial interactions may be less cost-effective than modeling spectral correlations with the same resources.  \textbf{Secondly}, the HSI representations are modulated by the  mask in the CASSI system. The original Transformer without sufficient guidance may easily attend to many low-fidelity and less informative  image  regions when calculating self-attention. This may degrade the model efficiency. \textbf{Thirdly}, when using the original global Transformer~\cite{dosovitskiy2020image}, the computational complexity is quadratic to the spatial size. This burden is nontrivial and sometimes unaffordable. When using the local window-based Transformer~\cite{liu2021swin}, the receptive fields of the MSA module are limited within the position-specific windows and some highly related tokens may be neglected.

To cope with the above problems, we propose a novel method, Mask-guided Spectral-wise Transformer (MST), for HSI reconstruction. \textbf{Firstly}, in Fig.~\ref{fig:illustration} (a), we observe that each spectral channel of HSIs captures an incomplete part of the same scene due to the limitation of the specific wavelength. This indicates that the HSI representations are similar and complementary along  the spectral dimension. Hence, we propose a Spectral-wise MSA (S-MSA) to capture the long-range inter-spectra dependencies. Specifically, S-MSA treats each spectral channel feature as a token and calculates the self-attention along the spectral dimension. \textbf{Secondly}, in Fig.~\ref{fig:illustration} (b), a mask is used in the CASSI system to modulate HSIs. The light transmittance of different positions on the mask varies significantly. This indicates that the fidelity of the modulated spectral information is position-sensitive.  Therefore, we exploit the mask as a key clue and  present a novel Mask-guided Mechanism (MM) that directs the S-MSA module to pay attention to the regions with high-fidelity spectral representations. Meanwhile, MM also alleviates the limitation of S-MSA in modeling the spatial correlations of HSI representations. \textbf{Finally}, with our proposed techniques, we establish a series of extremely efficient MST models that surpass state-of-the-art (SOTA) methods by a large margin, as illustrated in Fig.~\ref{fig:teaser}.

Our contributions can be summarized as follows:
\begin{itemize}
	\vspace{-2.25mm}
	\item We propose a new method, MST, for HSI reconstruction. To the best of our knowledge, it is the first attempt to explore the potential of Transformer in this task. 
	\vspace{-1.25mm}
	\item We present a novel self-attention, S-MSA, to capture the  inter-spectra similarity and dependencies of HSIs.
	\vspace{-1.25mm}
	\item We customize an MM that directs S-MSA to pay attention to regions with high-fidelity HSI representations.
	\vspace{-5.25mm}
	\item Our MST dramatically outperforms SOTA methods on all scenes in simulation while requiring much cheaper Params and FLOPS. Besides, MST yields more visually pleasant results in real-world HSI reconstruction. 
\end{itemize}

\section{Related Work}
\subsection{HSI Reconstruction}
Traditional HSI reconstruction  methods~\cite{sparse_1,sparse_2,sparse_3,desci,non_local_1,non_local_2,gap_tv,tra_rela_1,tra_rela_2} are mainly based on hand-crafted priors. For example, GAP-TV~\cite{gap_tv} introduces the total variation prior. 
DeSCI~\cite{desci} exploits the low-rank property and non-local self-similarity. However, these model-based methods achieve unsatisfactory performance and generality due to the poor representing capacity. Recently, deep CNNs have been applied to learn the end-to-end mapping function of HSI reconstruction~\cite{lambda,hssp,dnu,gsm,tsa_net,resu,self} to achieve promising performance. TSA-Net~\cite{tsa_net} uses three spatial-spectral self-attention modules to capture the dependencies in compressed spatial or spectral dimensions. The additional costs are nontrivial while the improvement is limited. 
DGSMP\cite{gsm} suggests an interpretable HSI restoration method with learned Gaussian Scale Mixture (GSM) prior. These CNN-based methods yield impressive performance but show limitations in modeling inter-spectra similarity and correlations. 
Besides, the \emph{guidance effect of the mask} is under-studied.

\begin{figure*}[t]
	\begin{center}
		\begin{tabular}[t]{c} \hspace{-3mm}
			\includegraphics[width=1.0\textwidth]{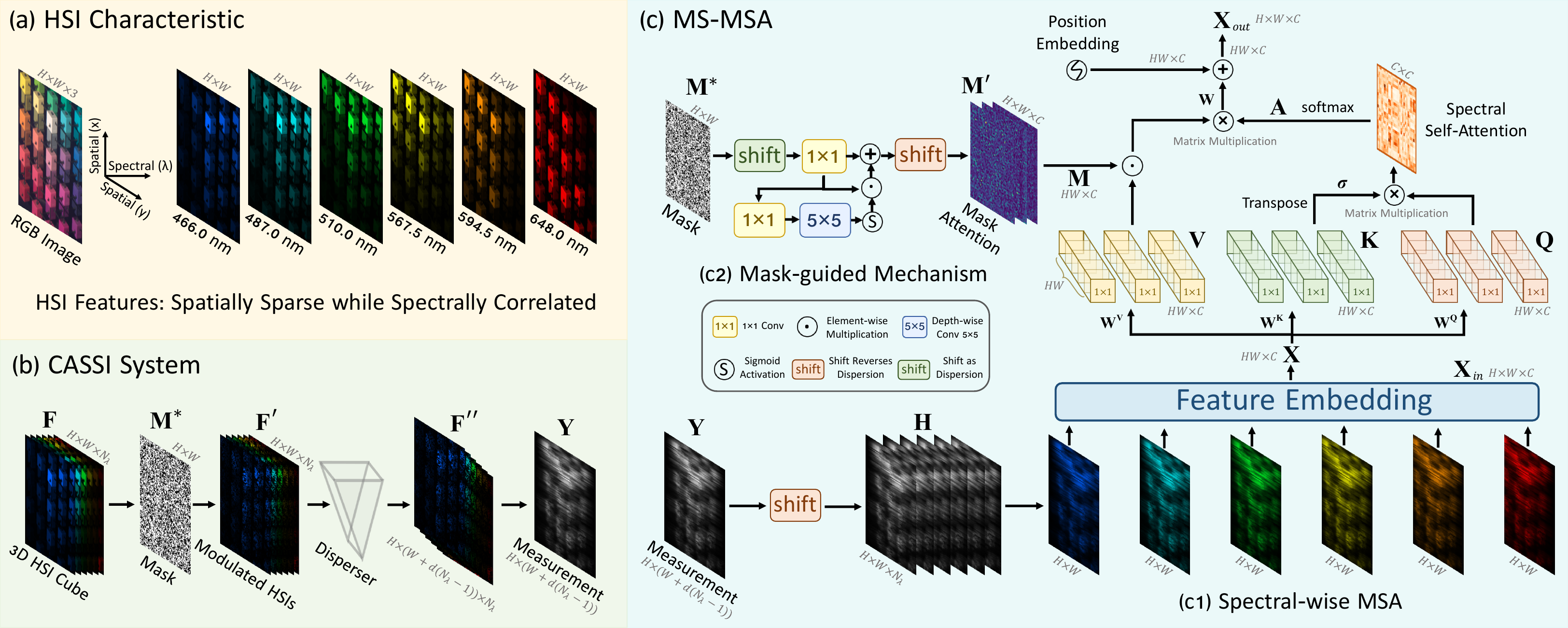}
		\end{tabular}
	\end{center}
	\vspace*{-6mm}
	\caption{\small Illustration of the proposed method. Our Mask-guided Spectral-wise Multi-head Self-Attention (MS-MSA) is motivated by the HSI characteristics and CASSI system. (a) The representations of HSIs are spatially sparse while spectrally correlated. (b) The CASSI system uses a mask to modulate the HSIs. (c) Our MS-MSA in stage 0 of MST. (c1) S-MSA  treats each spectral feature as a token and calculates self-attention along the spectral dimension. (c2) Mask-guided Mechanism  directs the Spectral-wise MSA to pay attention to spatial regions with high-fidelity HSI representations. Some components are omitted for simplification. Please refer to the text for details.}
	\label{fig:illustration}
	\vspace{-4mm}
\end{figure*}

\subsection{Vision Transformer}
Transformer is firstly proposed by~\cite{vaswani2017attention} for machine translation. 
Recently, Transformer has achieved great success in many high-level vision tasks, such as image classification~\cite{liu2021swin,arnab2021vivit,tc_1,tc_2,tc_3,xcit}, object detection~\cite{de_detr,to_1,liu2021swin,de_detr,DETR,dy_detr}, segmentation~\cite{tc_3,ts_1,liu2021swin,cao2021swin,SETR}, human pose estimation~\cite{tokenpose,transpose,prtr},  \emph{etc}. Due to its promising performance, Transformer has also been introduced into low-level vision~\cite{ipt,swinir,uformer,vsrt}.  SwinIR~\cite{swinir} uses Swin Transformer~\cite{swinir} blocks to build up a residual network and achieve SOTA results in image restoration. Chen \emph{et al.}~\cite{ipt} propose a large model IPT pre-trained on large-scale datasets with a multi-task learning scheme. However, these Transformers mainly aim to capture long-range dependencies of spatial regions. As for spectrally self-similar and mask-modulated HSIs, directly applying previous Transformers may be less effective in capturing spectral-wise correlations. Also, the MSA may pay attention to some  less informative spatial regions.

\begin{figure*}[t]
	\begin{center}
		\begin{tabular}[t]{c} \hspace{-3.4mm}
			\includegraphics[width=1.0\textwidth]{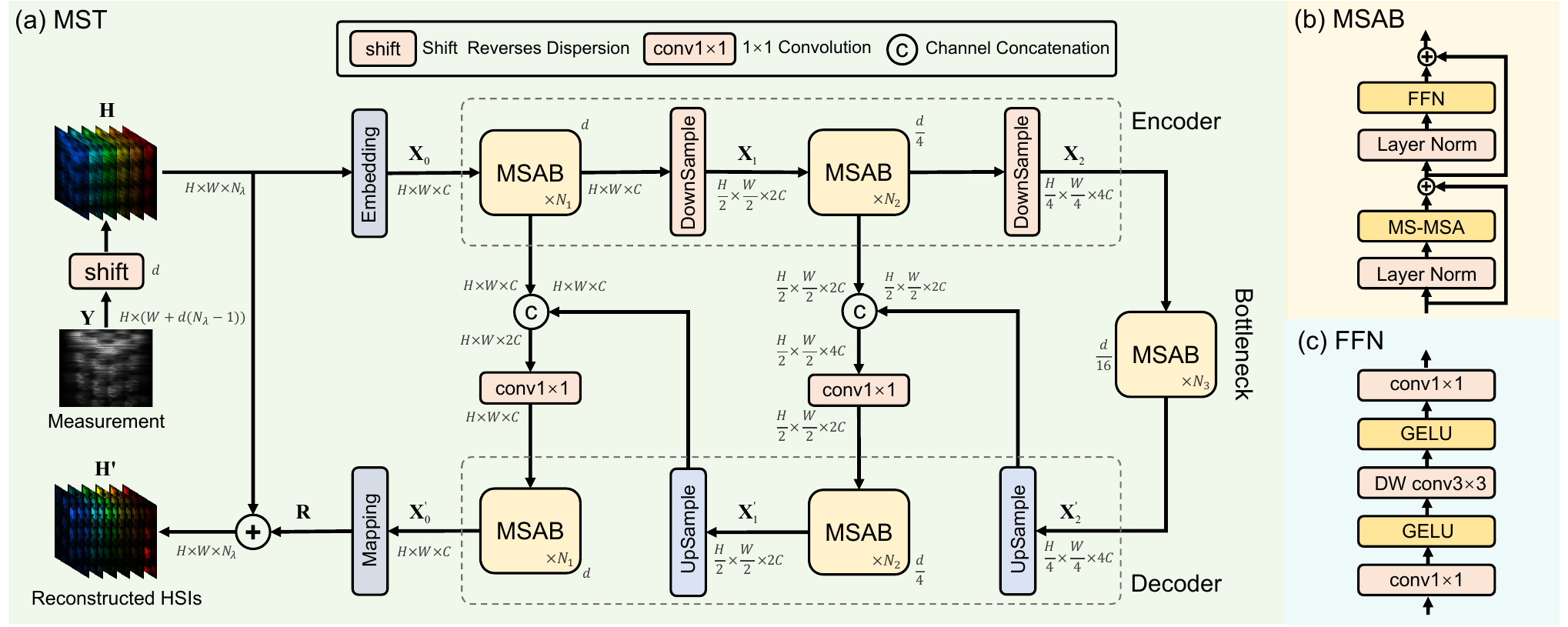}
		\end{tabular}
	\end{center}
	\vspace*{-6mm}
	\caption{\small The overall architecture of MST. (a) MST adopts a U-shaped structure that consists of an encoder, a bottleneck, and a decoder. (b) MSAB is composed of a Feed-Forward Network (FFN), an MS-MSA, and two layer normalization. (c) The components of FFN.}
	\label{fig:pipeline}
	\vspace{-4mm}
\end{figure*}

\vspace{-0.3mm}
\section{CASSI System}
\vspace{-0.3mm}
A concise CASSI principle is shown in Fig.~\ref{fig:illustration} (b). Given a 3D HSI cube, denoted by $\mathbf{F} \in \mathbb{R}^{H\times W \times N_{\lambda}}$, where $H$, $W$, and $N_\lambda$ represent the HSI's height, width, and number of wavelengths, respectively. $\mathbf{F}$ is firstly modulated by the coded aperture (physical mask) $\mathbf{M^*} \in \mathbb{R}^{H\times W}$ as 
\vspace{-1.2mm}
\begin{equation}\label{eq:Fprime}
\mathbf{F}' (:,:,n_{\lambda}) = \mathbf{F} (:,:,n_{\lambda}) \odot \mathbf{M^*},
\vspace{-1.2mm}
\end{equation} 
where $\mathbf{F}'$ denotes the modulated HSIs, $n_{\lambda} \in [1,\dots, N_{\lambda}]$ indexes the spectral channel, and $\odot$ denotes the element-wise multiplication. After passing through the disperser, $\mathbf{F}'$ becomes tilted and is considered to be  sheared along the $y$-axis. We use $\mathbf{F}'' \in \mathbb{R}^{H\times (W + d(N_{\lambda}-1)) \times N_{\lambda}}$ to denote the titled HSI cube, where $d$ represents the shifting step. We assume $\lambda_c$ to be the reference wavelength, \emph{i.e.}, $\mathbf{F}'' (:,:,n_{\lambda_c})$ is not sheared along the $y$-axis.  Then we have
\vspace{-1.2mm}
\begin{equation}\label{eq:Fprime_prime}
\mathbf{F}'' (u,v, n_{\lambda}) = \mathbf{F}'(x, y + d(\lambda_n - \lambda_c), n_{\lambda}),
\end{equation}
where $(u,v)$ represents the coordinate system on the detector plane, $\lambda_n$ denotes the wavelength of the $n_\lambda$-th channel, and $d(\lambda_n -\lambda_c)$ indicates the spatial shifting for the $n_\lambda$-th channel on $\mathbf{F}''$. Finally, the captured 2D compressed measurement $\mathbf{Y} \in \mathbb{R}^{H\times (W + d(N_{\lambda}-1))}$ can be obtained by
\vspace{-3.5mm}
\begin{equation}\label{eq:y-discret}
 \mathbf{Y} = \sum_{n_{\lambda}=1}^{N_{\lambda}}  \mathbf{F}'' (:,:, n_{\lambda}) +  \mathbf{G}, 
 \vspace{-1.5mm}
\end{equation}
where $\mathbf{G}\in \mathbb{R}^{H\times (W + d(N_{\lambda}-1))}$ is the imaging noise on the measurement, generated by the photon sensing detector. 

\vspace{-0.5mm}
\section{Method}
\vspace{-0.1mm}
\subsection{Overall Architecture}
\vspace{-0.5mm}
The overall architecture of MST is shown in Fig.~\ref{fig:pipeline} (a). We adopt a U-shaped structure that consists of an encoder, a bottleneck, and a decoder. MST is built up by Mask-guided Spectral-wise Attention Blocks (MSAB).  Firstly, we reverse the dispersion process (Eq.~\eqref{eq:Fprime_prime})  and shift back the measurement to obtain the initialized signal $\mathbf{H} \in \mathbb{R}^{H\times W \times N_{\lambda}}$ as
\vspace{-1.2mm}
\begin{equation}
\mathbf{H}(x,y,n_\lambda) = \mathbf{Y}(x,y-d(\lambda_n-\lambda_c)).
\vspace{-1.2mm}
\label{shift_back}
\end{equation} 
Then we feed $\mathbf{H}$ into the model.  \textbf{Firstly}, MST exploits a \emph{conv}3$\times$3 (convolution with kernel size = 3) layer to map $\mathbf{H}$ into feature $\mathbf{X}_0 \in  \mathbb{R}^{H\times W \times C}$. \textbf{Secondly}, $\mathbf{X}_0$ undergoes  $N_1$ MSABs, a downsample module, $N_2$ MSABs, and a downsample module to generate hierarchical features. The downsample module is a strided \emph{conv}4$\times$4 layer that downscales the feature maps and doubles the channels. Therefore, the feature of the $i$-th stage of the encoder is denoted as $\mathbf{X}_i \in \mathbb{R}^{\frac{H}{2^i} \times \frac{W}{2^i}  \times 2^{i}C}$. \textbf{Thirdly}, $\mathbf{X}_2$ passes through the bottleneck that consists of $N_3$ MSABs. \textbf{Subsequently}, We follow the spirit of U-Net~\cite{unet} and design a symmetrical structure as the decoder.  In particular, the upsample module is  a strided \emph{deconv}2$\times$2 layer. The skip connections are exploited for feature aggregation between the encoder and decoder to alleviate the information loss caused by the downsample operations. Similarly, the feature of the $i$-th stage of the decoder is denoted as $\mathbf{X}'_i \in \mathbb{R}^{\frac{H}{2^i} \times \frac{W}{2^i}  \times 2^{i}C}$. After passing through the decoder, the feature maps undergo a \emph{conv}3$\times$3 layer to generate the residual HSIs $\mathbf{R} \in \mathbb{R}^{H\times W \times N_{\lambda}}$. \textbf{Finally}, the reconstructed HSIs $\mathbf{H}' \in \mathbb{R}^{H\times W \times N_{\lambda}}$  can be obtained by the sum of $\mathbf{R}$ and $\mathbf{H}$, \emph{i.e.}, $\mathbf{H}' = \mathbf{H} + \mathbf{R}$.

In implementation, we set $C$ to 28 and change the combination $(N_1,N_2,N_3)$ to establish a series of MST models with small, medium, and large model sizes and computation costs: MST-S (2,2,2), MST-M (2,4,4), and MST-L (4,7,5).

The basic unit of MST is MSAB. As shown in Fig.~\ref{fig:pipeline} (b), MSAB consists of two layer normalization, a Mask-guided Spectral-wise MSA (MS-MSA), and a Feed-Forward Network (FFN). The details of FFN are depicted in Fig.~\ref{fig:pipeline} (c).

\subsection{Spectral-wise Multi-head Self-Attention}
The non-local self-similarity is often exploited in HSI reconstruction but is not well modeled by CNN-based methods. Due to the effectiveness of Transformer in capturing non-local long-range dependencies and its impressive performance in other vision tasks, we aim to explore the potential of Transformer in HSI reconstruction. However, there are two main issues when directly applying Transformer to HSI restoration. The first problem is that original Transformers model long-range dependencies in spatial dimensions. But the HSI representations are spatially sparse and spectrally  correlated, as shown in Fig.~\ref{fig:illustration} (a). Capturing spatial-wise interactions may be less cost-effective than modeling spectral-wise correlations.  Hence, we propose S-MSA that treats each spectral feature map as a token and calculates self-attention along the spectral dimension. Fig.~\ref{fig:illustration} (c1) shows the S-MSA used in stage 0 of MST. The input $\mathbf{X}_{in} \in \mathbb{R}^{H\times W \times C}$ is reshaped into tokens $\mathbf{X} \in \mathbb{R}^{HW \times C}$. Then $\mathbf{X}$ is linearly projected into \emph{query} $\mathbf{Q} \in \mathbb{R}^{HW \times C}$, \emph{key} $\mathbf{K} \in \mathbb{R}^{HW \times C}$, and \emph{value} $\mathbf{V} \in \mathbb{R}^{HW \times C}$: 
\begin{equation}
\mathbf{Q} = \mathbf{X}\mathbf{W}^\mathbf{Q}, \mathbf{K} = \mathbf{X}\mathbf{W}^\mathbf{K}, \mathbf{V} = \mathbf{X}\mathbf{W}^\mathbf{V},
\label{linear_proj}
\end{equation}
where $\mathbf{W}^\mathbf{Q}$, $\mathbf{W}^\mathbf{K}$, and $\mathbf{W}^\mathbf{V} \in \mathbb{R}^{C \times C}$ are learnable parameters; $biases$ are omitted for simplification. Subsequently, we respectively split $\mathbf{Q}$, $\mathbf{K}$, and $\mathbf{V}$ into $N$ \emph{heads} along the spectral channel dimension: $\mathbf{Q} = [\mathbf{Q}_1,\ldots,\mathbf{Q}_N]$, $\mathbf{K} = [\mathbf{K}_1,\ldots,\mathbf{K}_N]$ and $\mathbf{V} = [\mathbf{V}_1,\ldots,\mathbf{V}_N]$. The dimension of each head is $d_h = \frac{C}{N}$. Different from original MSAs, our S-MSA treats each spectral representation as a token and calculates the self-attention for each $head_j$:
\begin{equation}
	\mathbf{A}_j = \text{softmax}(\sigma_j \mathbf{K}_j^\text{T} \mathbf{Q}_j), ~~{head}_j  =\mathbf{V}_j \mathbf{A}_j, 
	\label{s-attention}
\end{equation}
where $\mathbf{K}_j^\text{T}$ denotes the transposed matrix of $\mathbf{K}_j$. We note that the spectral density varies significantly with respect to the wavelengths. Thus, we exploit a learnable parameter $\sigma_j \in \mathbb{R}^{1}$ to adapt the self-attention $\mathbf{A}_j$ by re-weighting the matrix multiplication $\mathbf{K}_j^\text{T} \mathbf{Q}_j$ inside $head_j$.  Please note that Fig.~\ref{fig:illustration} (c1) depicts the situation with $N$ = 1 and some details are omitted for simplification. Subsequently, the outputs of $N$ \emph{heads} are concatenated in spectral wise to undergo a linear projection and then is added with a position embedding:
\begin{equation}
\text{S-MSA}(\mathbf{X}) =\big(\mathop{\text{Concat}}\limits_{j=1}^{N}(head_{j})\big)\mathbf{W} + f_p(\mathbf{V}),
\label{agg_heads}
\end{equation}
where $\mathbf{W} \in \mathbb{R}^{C \times C}$ are learnable parameters, $f_p(\cdot)$ is the function to generate position embedding. It consists of two depth-wise \emph{conv}3$\times$3 layers, a GELU activation, and reshape operations. The HSIs are sorted by the wavelength along the spectral dimension. Therefore, we exploit this embedding to encode the position information of different spectral channels. Finally, we reshape the result of Eq.~\eqref{agg_heads} to obtain the output feature map $\mathbf{X}_{out} \in \mathbb{R}^{H\times W \times C}$.

We analyze the computational complexity of S-MSA and compare it with other MSAs. We only compare the  difference, \emph{i.e.}, the  self-attention mechanism in Eq.~\eqref{s-attention}:
\vspace{-1mm}
\begin{equation}
\begin{aligned}
O(\text{S-MSA}) = \frac{2HWC^2}{N},~O(\text{G-MSA}) = 2(HW)^2C,\\
O(\text{W-MSA}) = 2(M^2)^2(\frac{HW}{M^2})C = 2M^2HWC,
\end{aligned}
\vspace{-1mm}
\end{equation}
where G-MSA denotes the original global MSA~\cite{dosovitskiy2020image}, W-MSA denotes the local window-based MSA~\cite{liu2021swin}, and $M$ represents the window size. The computational complexity of S-MSA and W-MSA is linear to the spatial size $HW$. This cost is much cheaper than that of G-MSA (quadratic to $HW$). Meanwhile, S-MSA treats a whole spectral feature map as a token. Thus, the receptive field of our S-MSA is global and not limited to position-specific windows when compared to that of W-MSA. Please refer to the supplementary material (SM) for detailed analyses and comparisons.

\vspace{-0.5mm}
\subsection{Mask-guided Mechanism}
\vspace{-0.5mm}
\label{sec:mask}
The second problem of directly using Transformer for HSI restoration is that original Transformers may attend to some less informative spatial regions with low-fidelity HSI representations. In CASSI, a physical mask is used to modulate the HSIs. Thus, the light transmittance of different positions on the mask varies. As a result, the fidelity of the modulated spectral information is position-sensitive. This observation motivates us  that the mask should be used as a clue to direct the model to attend to regions with high-fidelity HSI representations. In this part, we firstly analyze the usage of the mask in  previous CNN-based methods, and then introduce our Mask-guided Mechanism (MM).

\noindent\textbf{Previous Mask Usage Scheme.}  Previous CNN-based methods~\cite{tsa_net,resu,gapnet} conduct an inner product between the initialized HSIs $\mathbf{H}$ and the mask $\mathbf{M}^*$ to generate a modulated input. This scheme introduces spatial fidelity information but suffers from the following limitations: \textbf{(i)} This operation corrupts the input HSI representations, causes the information loss, and leads to spatial discontinuity. \textbf{(ii)} This scheme only operates at the input. The \emph{guidance effect of the mask} in directing the network to pay attention to regions with high-fidelity HSI representations is not fully explored. \textbf{(iii)} This scheme does not exploit learnable parameters to adaptively model the spatial-wise correlations.

\noindent\textbf{Our MM.}  Different from previous methods, our MM preserves all the input HSI representations and learns to direct S-MSA to pay attention to the spatial regions with high-fidelity spectral representations. To be specific, given the mask $\mathbf{M}^* \in \mathbb{R}^{H\times W}$ shown in Fig.~\ref{fig:illustration} (c2), since the modulated HSIs are shifted by the disperser of the CASSI system, we firstly shift $\mathbf{M}^*$ like the dispersion process:
\vspace{-0.5mm}
\begin{equation}
\mathbf{M}_s(x,y,n_\lambda) = \mathbf{M}^*(x,y+d(\lambda_n-\lambda_c)),
\vspace{-0.5mm}
\label{shift}
\end{equation}
where $\mathbf{M}_s\in\mathbb{R}^{H\times (W+d(N_{\lambda}-1))\times N_{\lambda}}$ denotes the shifted version of $\mathbf{M}^*$. The shifted regions out of the range in $y$-axis on $\mathbf{M}^*$ are set to 0. Please note that  Fig.~\ref{fig:illustration} (c2) shows the MM used in stage 0 of MST. To match the scale of the feature maps in stage $i$ of MST, $\mathbf{M}_s$ needs to pass through the same downsample operations in Fig.~\ref{fig:pipeline} (a). Subsequently, $\mathbf{M}_s$ undergoes a \emph{conv}1$\times$1 layer and then is input to two paths. The upper path is an identity mapping to retain the original fidelity information. The lower path undergoes a \emph{conv}1$\times$1 layer, a depth-wise  \emph{conv}5$\times$5 layer, a sigmoid activation, and an inner product with the top features. S-MSA is effective in capturing inter-spectra dependencies but shows limitations in modeling spatial interactions of HSI representations. Thus, the lower path is designed to capture the spatial-wise correlations. Then we have
\begin{equation}
\begin{aligned}
\mathbf{M}'_s = (\mathbf{W}_1\mathbf{M}_s)\odot(1+\delta(f_{dw}(\mathbf{W}_2\mathbf{W}_1\mathbf{M}_s)),
\end{aligned}
\end{equation}
where $\mathbf{W}_1$ and $\mathbf{W}_2$ are the learnable parameters of the two \emph{conv}1$\times$1 layers, $f_{dw}(\cdot)$ denotes the mapping function of the depth-wise \emph{conv}5$\times$5 layer, $\delta(\cdot)$ represents the sigmoid activation, and $\mathbf{M}'_s \in \mathbb{R}^{H\times (W+d(N_{\lambda}-1))\times C}$ denotes the intermediate feature maps. To spatially align the mask attention map with the modulated HSIs $\mathbf{F}'$ in the CASSI system (Fig.~\ref{fig:illustration} (b)) and the initialized input $\mathbf{H}$ of MST (Fig.~\ref{fig:pipeline} (a)), we reverse the dispersion process and shift back $\mathbf{M}'_s$ to obtain the mask attention map $\mathbf{M}' \in \mathbb{R}^{H\times W \times C}$ as
\begin{equation}
\mathbf{M}'(x,y,n_\lambda) = \mathbf{M}'_s(x,y-d(\lambda_n-\lambda_c),n_\lambda),
\label{shift_back_mask}
\end{equation}
where we change the range of channel from 1 $\sim N_\lambda$ to 1 $\sim C$ to match the dimensions of $\mathbf{M}'_s$. We reshape $\mathbf{M}'$ into $\mathbf{M} \in \mathbb{R}^{HW \times C}$ to match the dimensions of $\mathbf{V}$. Then we split $\mathbf{M}$ into $N$ \emph{heads} in spectral wise: $\mathbf{M} = [\mathbf{M}_1,\ldots,\mathbf{M}_N]$. For each $head_j$, MM conducts its guidance by re-weighting $\mathbf{V}_j$ using $\mathbf{M}_j$. Hence, the S-MSA module just needs to make a simple modification by re-writing $head_j$ in Eq.~\eqref{s-attention}:
\begin{equation}
{head}_j  = (\mathbf{M}_j \odot \mathbf{V}_j) \mathbf{A}_j, 
\label{s-attention_mask}
\end{equation}
where $\mathbf{M}_j \in \mathbb{R}^{HW \times d_h}$, and $\odot$ denotes the element-wise multiplication. The subsequent steps (Eq.~\eqref{agg_heads}) of S-MSA remain unchanged when using MM to direct S-MSA.

\begin{table*}[t]
	\centering
	\resizebox{0.98\textwidth}{!}{
		\setlength{\tabcolsep}{0.7mm}
		\centering
		\begin{tabular}{c|cccccccccccccccccc|cccccc} 
			\toprule[0.15em]
			& \multicolumn{2}{c}{TwIST~\cite{twist}} & \multicolumn{2}{c}{GAP-TV~\cite{gap_tv}} & \multicolumn{2}{c}{DeSCI~\cite{desci}} 
			& \multicolumn{2}{c}{$\lambda$-net~\cite{lambda}} & \multicolumn{2}{c}{HSSP~\cite{hssp}}  &\multicolumn{2}{c}{DNU~\cite{dnu}} &\multicolumn{2}{c}{DIP-HSI~\cite{self}}  & \multicolumn{2}{c}{TSA-Net~\cite{tsa_net}} & \multicolumn{2}{c|}{~DGSMP~\cite{gsm}~~}& \multicolumn{2}{c}{~\bf MST-S} & \multicolumn{2}{c}{\bf MST-M} & \multicolumn{2}{c}{\bf MST-L}\\
			Scene~&~PSNR~ & SSIM & PSNR & SSIM & PSNR & SSIM & PSNR & SSIM & PSNR & SSIM & PSNR & SSIM & PSNR & SSIM & PSNR & SSIM & PSNR & ~SSIM ~& ~PSNR~ & SSIM & PSNR & SSIM &PSNR &SSIM\\
			\midrule[0.15em]
			1  & 25.16 & 0.700 & 26.82 & 0.754 & 27.13 & 0.748 & 30.10 & 0.849 & 31.48 & 0.858 & 31.72 & 0.863 &32.68 &0.890 & 32.03 & 0.892 &33.26 &0.915  & 34.71 & 0.930 & 35.15 & 0.937 & {\bf35.40} &{\bf0.941} \\
			2  & 23.02 & 0.604 & 22.89 & 0.610 & 23.04 & 0.620 & 28.49 & 0.805 & 31.09 & 0.842 &31.13 & 0.846 &27.26 &0.833 & 31.00 & 0.858 & 32.09 &0.898 & 34.45 & 0.925 & 35.19 & 0.935 &\bf 35.87 &\bf 0.944\\
			3  & 21.40 & 0.711 & 26.31 & 0.802 & 26.62 & 0.818 & 27.73 & 0.870 & 28.96 & 0.823 & 29.99 & 0.845 &31.30 &0.914 & 32.25 & 0.915 &33.06 &0.925 & 35.32 & 0.943 & 36.26 & 0.950 & {\bf36.51} &{\bf0.953}\\
			4  & 30.19 & 0.851 & 30.65 & 0.852 & 34.96 & 0.897 & 37.01 & 0.934 & 34.56 & 0.902 & 35.34 & 0.908 &40.54 &0.962 & 39.19 & 0.953 &40.54 &0.964  &41.50 & 0.967 & \bf 42.48 & \bf 0.973 & {42.27} &{\bf0.973}\\
			5  & 21.41 & 0.635 & 23.64 & 0.703 & 23.94 & 0.706 & 26.19 & 0.817 & 28.53 & 0.808 & 29.03 & 0.833 &29.79 &0.900 & 29.39 & 0.884 &28.86 &0.882  & 31.90 & 0.933 & 32.49 & 0.943 & {\bf32.77} &{\bf0.947}\\
			6  & 20.95 & 0.644 & 21.85 & 0.663 & 22.38 & 0.683 & 28.64 & 0.853 & 30.83 & 0.877 & 30.87 & 0.887 &30.39 &0.877 & 31.44 & 0.908 &33.08 &0.937 & 33.85 & 0.943 & 34.28 & 0.948 & {\bf34.80} &{\bf0.955}\\
			7  & 22.20 & 0.643 & 23.76 & 0.688 & 24.45 & 0.743 & 26.47 & 0.806 & 28.71 & 0.824 & 28.99 & 0.839 &28.18 &0.913 & 30.32 & 0.878 & 30.74 &0.886  & 32.69 & 0.911 & 33.29 & 0.921 & \bf 33.66 &{\bf0.925}\\
			8  & 21.82 & 0.650 & 21.98 & 0.655 & 22.03 & 0.673 & 26.09 & 0.831 & 30.09 & 0.881 & 30.13 & 0.885 &29.44 &0.874 &29.35 & 0.888 &31.55 &0.923 & 31.69 & 0.933 & 32.40 & 0.943 & {\bf32.67} &{\bf0.948}\\
			9  & 22.42 & 0.690 & 22.63 & 0.682 & 24.56 & 0.732 & 27.50 & 0.826 & 30.43 & 0.868 & 31.03 & 0.876 &34.51 &0.927 & 30.01 & 0.890 &31.66 &0.911  & 34.67 & 0.939 & 35.35 & 0.942 & {\bf35.39} &{\bf0.949}\\
			10 & 22.67 & 0.569 & 23.10 & 0.584 & 23.59  & 0.587 & 27.13 & 0.816 & 28.78 & 0.842 & 29.14 & 0.849 &28.51 &0.851 &29.59 & 0.874 &31.44 &0.925 &31.82 & 0.926 & 32.53 & 0.935 & {\bf32.50} &{\bf0.941}\\
			\midrule
			Avg  & 23.12 & 0.669 & 24.36 & 0.669 & 25.27 & 0.721 & 28.53 & 0.841 & 30.35 & 0.852 & 30.74 & 0.863 &31.26 &0.894 &31.46 & 0.894 &32.63 &0.917 & 34.26 & 0.935 & 34.94 &0.943 & {\bf35.18} &{\bf0.948}\\
			\bottomrule[0.15em]
	\end{tabular}}
	\vspace{-2mm}
	\caption{\small Quantitative results on 10 scenes in simulation. PSNR and SSIM are reported. MSTs significantly surpass other competitors.}\label{Tab:performance}
	\vspace{-4mm}
\end{table*}

\begin{figure*}[t]
	\begin{center}
		\begin{tabular}[t]{c} \hspace{-4.8mm}
			\includegraphics[width=1.0\textwidth]{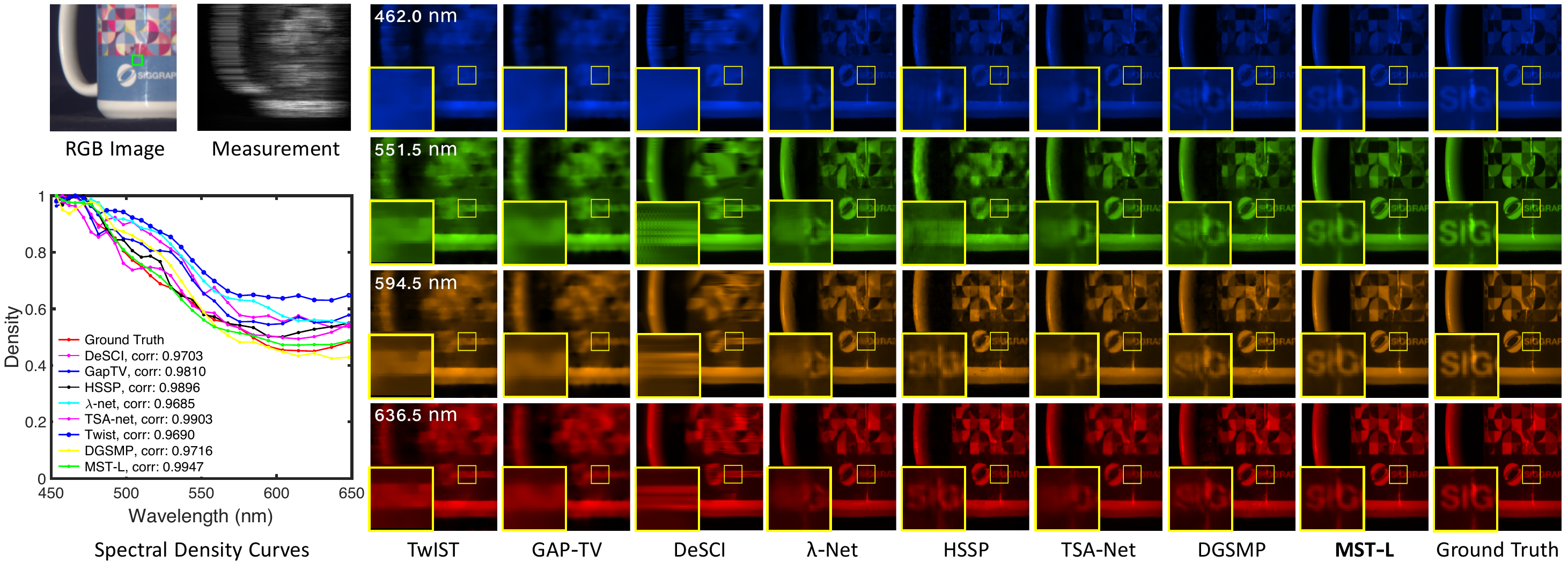}
		\end{tabular}
	\end{center}
	\vspace*{-8mm}
	\caption{\small Reconstructed simulation HSI comparisons of  \emph{Scene} 5 with 4 out of 28 spectral channels. 7 SOTA algorithms and our MST-L are included. The spectral curves (bottom-left) are corresponding to the selected green box of the RGB image. Zoom in for a better view. }
	\label{fig:simulation}
	\vspace{-4.5mm}
\end{figure*}

\vspace{-1.5mm}
\section{Experiments}
\vspace{-0.5mm}
\subsection{Experimental Settings}
\vspace{-0.5mm}
Following the settings of TSA-Net~\cite{tsa_net}, we adopt 28 wavelengths from 450 nm to 650 nm derived by spectral interpolation manipulation for HSIs. We perform experiments on both simulation and real HSI datasets.

\noindent \textbf{Simulation HSI Data.} We use two simulation hyperspectral image datasets, CAVE~\cite{cave} and KAIST~\cite{kaist}. CAVE dataset is composed of 32 hyperspectral images at a spatial size of 512$\times$512. KAIST dataset consists of 30 hyperspectral images at a spatial size of 2704$\times$3376. Following  the schedule of  TSA-Net~\cite{tsa_net}, we adopt CAVE as the training set. 10 scenes from KAIST are  selected for testing. 

\noindent \textbf{Real HSI Data.} We use the real HSI dataset   collected by the CASSI system developed in TSA-Net~\cite{tsa_net}.

\noindent \textbf{Evaluation Metrics.} We adopt peak signal-to-noise ratio (PSNR) and structural similarity (SSIM)~\cite{ssim} as the metrics to evaluate the HSI  reconstruction performance.

\noindent \textbf{Implementation Details.} We implement MST in Pytorch. All the models are trained with Adam~\cite{adam} optimizer ($\beta_1$ = 0.9 and $\beta_2$ = 0.999) for 300 epochs. The learning rate is set to 4$\times$10$^{-4}$ in the beginning and is halved every 50 epochs during the training procedure. When conducting experiments on simulation data, patches at a spatial size of 256$\times$256 cropped from the 3D cubes are fed into the networks. As for real hyperspectral image reconstruction, the patch size is set to 660$\times$660 to match the real-world measurement. The shifting step $d$ in dispersion is set to 2. Thus, the measurement sizes are 256$\times$310 and 660$\times$714 for synthetic and real HSI datasets. The shifting step in reversed dispersion is ${d}/{4^i}, i = 0,1,2$ for the $i$-th stage of MST. The batch size is 5. Random flipping and rotation are used for data augmentation. The models are trained on one RTX 8000 GPU. The training objective is to minimize the Root Mean Square Error (RMSE) and Spectrum Constancy Loss~\cite{Zhao_2019_CVPR} between the reconstructed and ground-truth HSIs.
\begin{figure*}[t]
	\begin{center}
		\begin{tabular}[t]{c} \hspace{-2.7mm} 
			\includegraphics[width=0.98\textwidth]{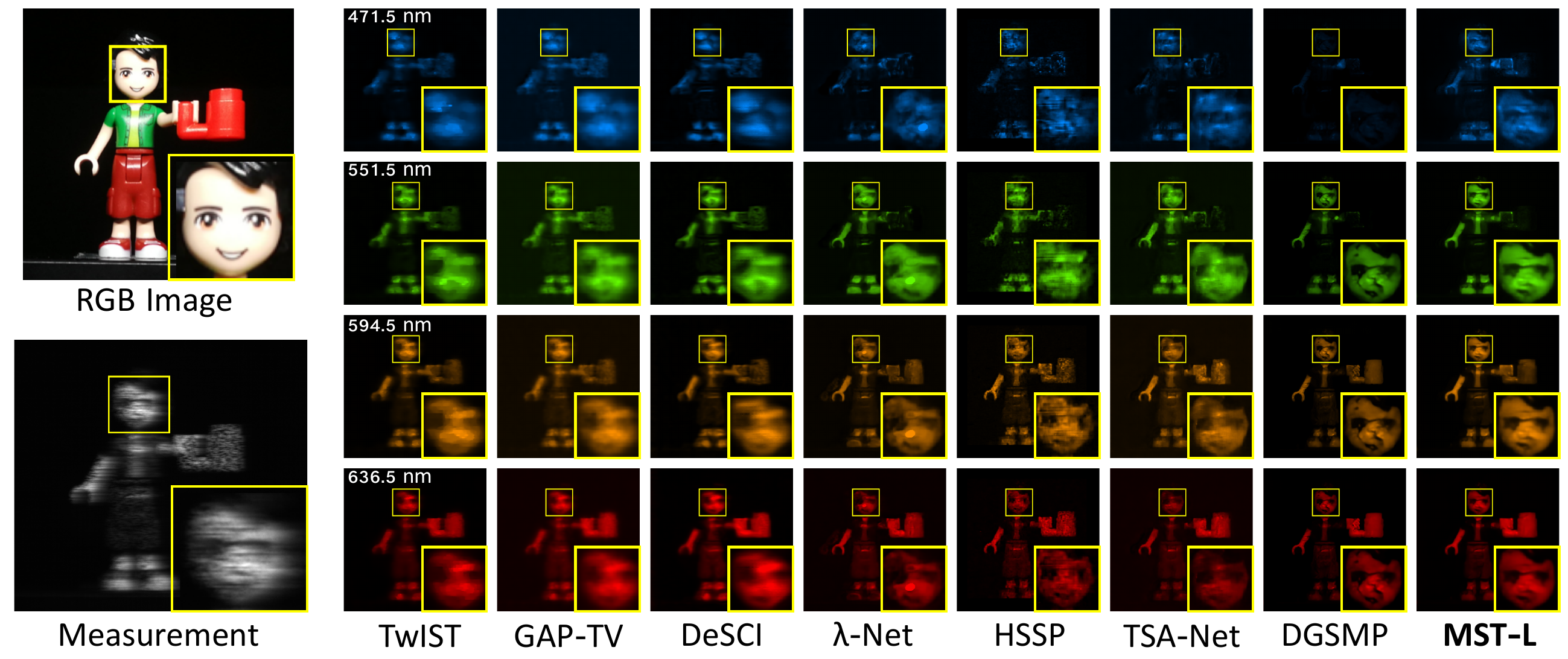}
		\end{tabular}
	\end{center}
	\vspace*{-7mm}
	\caption{\small Reconstructed real HSI comparisons of  \emph{Scene} 3 with 4 out of 28 spectral channels. Seven SOTA algorithms and our MST-L are included. MST-L reconstructs more detailed contents and suppresses more noise. Please zoom in for better visualization performance. }
	\label{fig:real}
	\vspace{-5mm}
\end{figure*}

\vspace{-4.5mm}
\subsection{Quantitative Results}
\vspace{-0.5mm}
We compare our MST with several SOTA HSI reconstruction algorithms, including three model-based methods (TwIST~\cite{twist}, GAP-TV~\cite{gap_tv}, and DeSCI~\cite{desci}) and six CNN-based methods ($\lambda$-net~\cite{lambda}, HSSP~\cite{hssp}, DNU~\cite{dnu}, PnP-DIP-HSI~\cite{self}, TSA-Net~\cite{tsa_net}, and DGSMP~\cite{gsm}). For fair comparisons, all methods are tested with the same settings as DGSMP~\cite{gsm}. The PSNR and SSIM results of different methods  on 10 scenes in the simulation datasets are listed in Tab.~\ref{Tab:performance}. The Params and FLOPS (test size = 256$\times$256) of \textbf{open-source} CNN-based algorithms are reported in Tab.~\ref{tab:model_size}. It can be observed from these two tables that our MSTs significantly surpass previous methods by a large margin on all the 10 scenes while requiring much cheaper memory and computational costs. More specifically, our best model, MST-L surpasses DGSMP, TSA-Net, and $\lambda$-net by 2.55, 3.72, and 6.65 dB while costing 54.0\% (2.03 / 3.76), 4.6\%, and 3.2\% Params and 4.4\% (28.15 / 646.65), 25.6\%, and 23.9\% FLOPS. Surprisingly, even our smallest model, MST-S outperforms DGSMP, TSA-Net, PnP-DIP-HSI, DNU, and $\lambda$-net by 1.63, 2.80, 3.00, 3.52, and 5.73 dB while requiring 24.7\%, 2.1\%, 2.7\%, 78.2\%, and 1.5\% Params and 2.0\%, 11.8\%, 20.1\%, 7.9\%, and 11.0\% FLOPS. 

To intuitively show the superiority of our MST, we provide PSNR-Params-FLOPS comparisons of different reconstruction algorithms in Fig.~\ref{fig:teaser}. The vertical axis is PSNR (performance), the horizontal axis is FLOPS (computational cost), and the circle radius is Params (memory cost). It can be seen that our MSTs take up the upper-left corner, exhibiting the extreme efficiency advantages of our method.

\vspace{-1.8mm}
\subsection{Qualitative Results}
\vspace{-1.2mm}
\noindent\textbf{Simulation HSI Reconstruction.} Fig.~\ref{fig:simulation} visualizes the reconstructed simulation HSIs of \emph{Scene} 5 with 4 out of 28 spectral channels using seven SOTA methods and our MST-L. Please zoom in for a better view. As can be seen from the reconstructed HSIs (right) and the zoom-in patches of the selected yellow boxes, previous methods are less favorable to restore HSI details. They either yield over-smooth results sacrificing fine-grained structural contents and textural details, or introduce undesirable chromatic artifacts and blotchy textures. In contrast, MST-L is more capable of recovering textural details and fine-grained contents, and preserving  the spatial smoothness of the homogeneous regions. This is mainly because our MST-L enjoys the guidance of modulation information and captures the long-range dependencies of different spectral channels. In addition, we plot the spectral density curves (bottom-left) corresponding to the picked region of the green box in the RGB image (top-left). The highest correlation and coincidence between our curve and the ground truth demonstrate the spectral-wise consistency restoration effectiveness of our MST.

\noindent\textbf{Real HSI Reconstruction.} We further apply our proposed approach to real HSI reconstruction. Similar to~\cite{tsa_net,gsm}, we re-train our model (MST-L) on all scenes of CAVE~\cite{cave} and KAIST~\cite{kaist} datasets. To simulate real imaging situations, we inject 11-bit shot noise into the measurements during training. Visual comparisons are shown in Fig.~\ref{fig:real}. Our MST-L surpasses previous algorithms in terms of high-frequency structural detail reconstruction and real noise suppression. These results clearly suggest the HSI reconstruction superiority and generalization ability of our proposed method. Please refer to the SM for more qualitative comparisons.

\begin{table*}[t]\vspace{-3mm}
	\subfloat[\small Break-down ablation study towards higher performance. \label{tab:breakdown}]{ 
		\scalebox{0.78}{
			\begin{tabular}{c c c  c c c c}
				\toprule
				\rowcolor{color3} Baseline &S-MSA &MM   &PSNR &SSIM &Params (M) &FLOPS (G) \\
				\midrule
				\checkmark & & &32.29 &0.897 &0.53 &7.43 \\
				\checkmark  &\checkmark & &33.18 &0.923 &0.70 &10.36 \\
				\checkmark  &\checkmark &\checkmark &\bf 34.26 &\bf 0.935 &0.93 &12.96 \\
				\bottomrule
	\end{tabular}}}\hspace{4mm}\vspace{0mm}
	\subfloat[\small Ablation study of different self-attention mechanisms.\label{tab:attention}]{
		\scalebox{0.65}{
			\begin{tabular}{l c c c c c}
				\toprule
				\rowcolor{color3} Method &Baseline &Global MSA &Local W-MSA & Swin W-MSA &\bf S-MSA\\
				\midrule
				PSNR &32.29 &32.67 &32.75 &32.86 &\bf 33.18  \\
				SSIM &0.897 &0.912 &0.916 &0.919 &\bf 0.923  \\
				Params (M)  &0.53 &0.70 &0.70 &0.70  &0.70  \\
				FLOPS (G)  &7.43 &11.88 &11.07 &11.07 &10.36 \\
				\bottomrule
	\end{tabular}}}\vspace{0mm}
	\subfloat[\small Params and FLOPS (test size = 256$\times$256) of different \textbf{open-source} CNN-based methods. \label{tab:model_size}]{
		\scalebox{0.68}{\begin{tabular}{l c c c c c c c c c}
				\toprule
				\rowcolor{color3}Method &{$\lambda$-net~\cite{lambda}}  &{DNU~\cite{dnu}}  &DIP-HSI~\cite{self} &{TSA-Net~\cite{tsa_net}}  &{DGSMP~\cite{gsm}}& {\bf MST-S} & {\bf MST-M} & {\bf MST-L}   \\
				\midrule
				PSNR &28.53   &30.74 &31.26  &31.46  &32.63 &\bf 34.26 &\bf 34.94 &\bf 35.18 \\
				SSIM  &0.841    &0.863  &0.894 &0.894  &0.917 &\bf 0.935 &\bf 0.943 &\bf 0.948 \\
				Params (M)  &62.64    &1.19  &33.85  &44.25  &3.76 &\bf 0.93 &\bf 1.50 &\bf 2.03\\
				FLOPS (G) &117.98    &163.48 &64.42   &110.06  &646.65 &\bf 12.96 &\bf 18.07 &\bf 28.15\\
				\bottomrule
	\end{tabular}}}\hspace{4mm}
	\subfloat[\small MM \emph{v.s} Previous usage of mask.\label{tab:mask}]{
		\scalebox{0.68}{
			\begin{tabular}{c c c c c}
				\toprule
				\rowcolor{color3} Method &Input &MM  &PSNR &SSIM   \\
				\midrule
				A  &$\mathbf{H}$ & &33.18 &0.923  \\
				B  &$\mathbf{H}\odot\mathbf{M}^*$ & &33.57 &0.927 \\
				\bf C &$\mathbf{H}$ &\checkmark &\bf 34.26 &\bf 0.935 \\
				D &$\mathbf{H}\odot\mathbf{M}^*$ &\checkmark &34.07 &0.932 \\
				\bottomrule
	\end{tabular}}}
	\vspace{-3mm}
	\caption{\small Ablations. We train models on CAVE~\cite{cave} and test on KAIST~\cite{kaist} in simulation. PSNR, SSIM, Params, and FLOPS are reported.}
	\label{tab:ablations}\vspace{-6mm}
\end{table*}

\vspace{-1.5mm}
\subsection{Ablation Study}
\vspace{-1.5mm}
In this part, we adopt the simulation HSI datasets~\cite{cave,kaist} to conduct ablation studies. The baseline model is derived by removing our S-MSA and MM from MST-S.

\noindent\textbf{Break-down Ablation.} We firstly perform a break-down ablation to investigate the effect of each component towards higher performance. The results are listed in Tab.~\ref{tab:breakdown}. The baseline model yields 32.29 dB. When we successively apply S-MSA and MM, the reconstruction performance continuously achieves 0.89 and 1.08 dB improvements. These results demonstrate the effectiveness of S-MSA and MM.

\begin{figure}[h]
	\begin{center}
		\begin{tabular}[t]{c} \hspace{-2.5mm} 
			\includegraphics[width=0.47\textwidth]{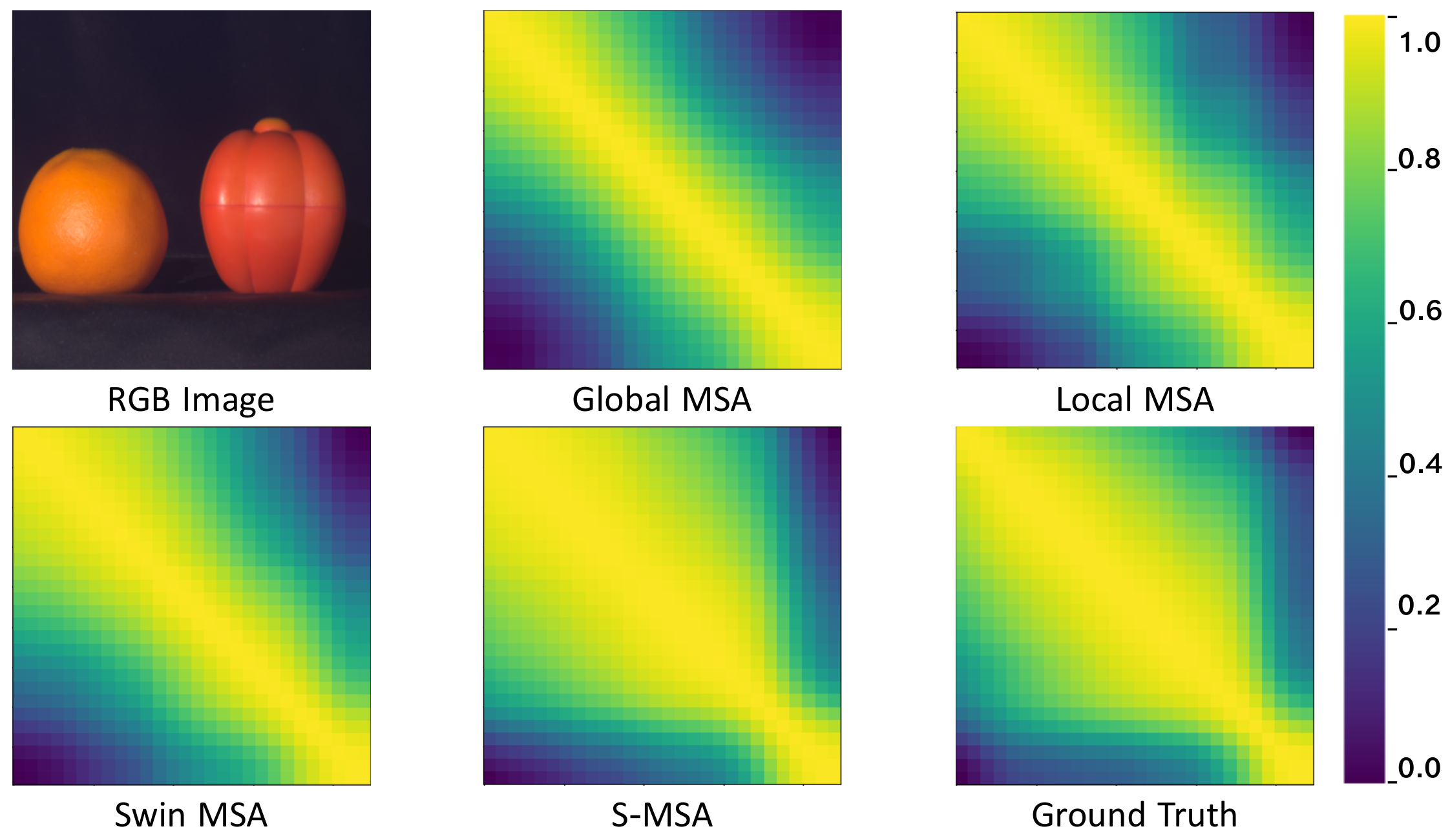}
		\end{tabular}
	\end{center}
	\vspace{-8mm}
	\caption{\small  Visualization of the correlation coefficients among spectral channels of HSIs reconstructed by models using different MSAs. The correlation coefficient map of the model equipped with our S-MSA is the most similar one to that of the ground truth. }
	\label{fig:sa_compare}
	\vspace{-5mm}
\end{figure}

\noindent\textbf{Self-Attention Comparison.} We compare  S-MSA with other self-attentions and report the  results in Tab.~\ref{tab:attention}. For fair comparisons, the Params of models using different self-attention schemes are set to the same value (0.70 M). The baseline model yields 32.29 dB while costing 0.53 M Params and 7.43 G FLOPS. We respectively apply the original global MSA~\cite{dosovitskiy2020image}, local window-based MSA~\cite{liu2021swin}, Swin W-MSA~\cite{liu2021swin}, and our S-MSA. The model gains by 0.38, 0.46, 0.57, and 0.89 dB while adding 4.45, 3.64, 3.64, and 2.93 G FLOPS. Our S-MSA yields the most significant improvement but requiries the least computational cost. Please note that we downscale the input feature of global MSA into $\frac{1}{4}$ size to avoid out of memory. We explain these results by the HSI characteristics  that the spectral representations are spatially sparse and spectrally highly self-similar. Hence, capturing spatial interactions may be less cost-effective than modeling inter-spectra dependencies. This evidence clearly verifies the efficiency superiority of our S-MSA. 

In addition, we further conduct visual analysis about different MSAs in Fig.~\ref{fig:sa_compare}. Specifically, we visualize the correlation coefficients between each spectral pair of HSIs reconstructed by models equipped with different MSAs. It can be observed that the correlation coefficient map of the model using our proposed S-MSA is the most similar one to that of the ground truth. These results demonstrate  the promising effectiveness of our S-MSA in modeling the inter-spectra similarity and  long-range spectral-wise dependencies.    

\begin{figure}[h]
	\begin{center}
		\begin{tabular}[t]{c} \hspace{-2.5mm} 
			\includegraphics[width=0.47\textwidth]{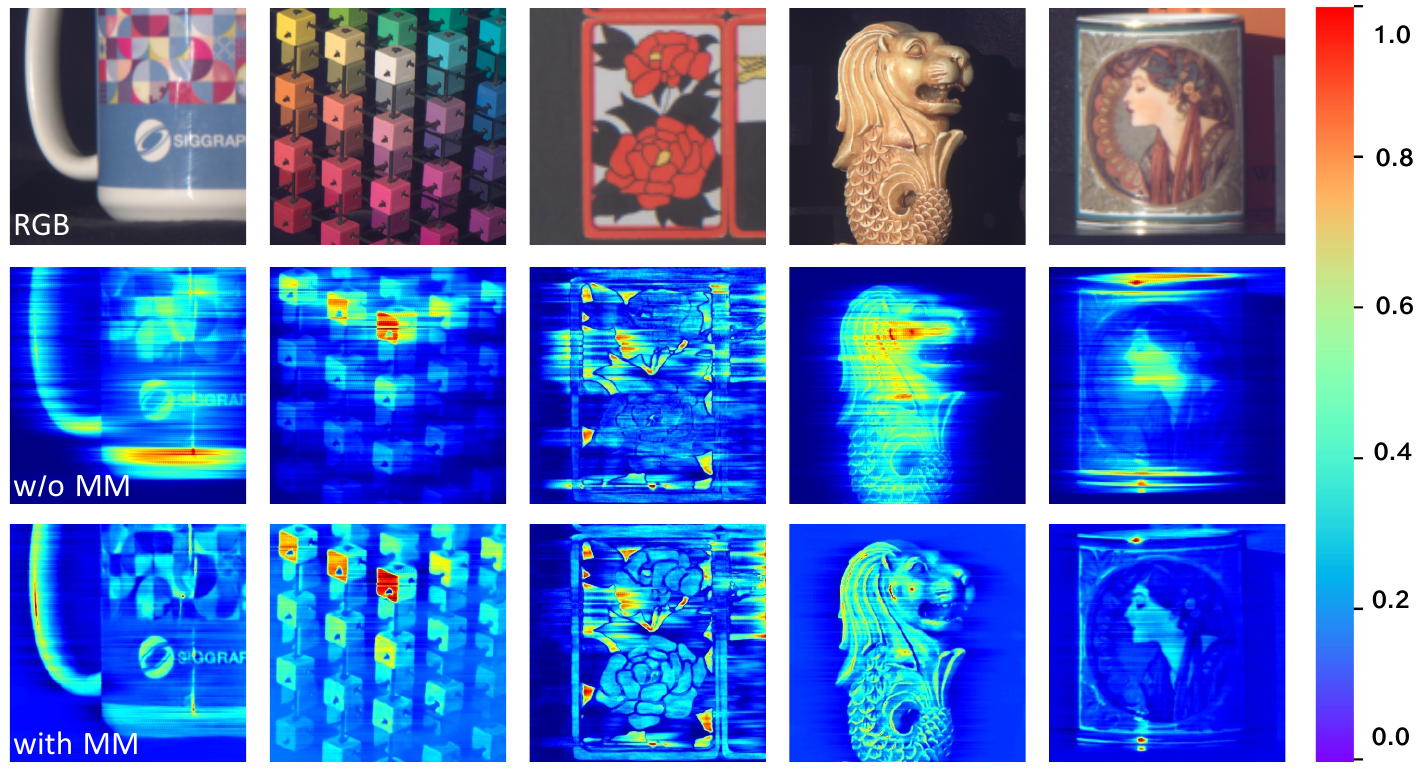}
		\end{tabular}
	\end{center}
	\vspace{-6mm}
	\caption{\small  Visual analysis of the feature maps of the last MSAB in MST-S. The top row shows the original RGB images. The middle and bottom rows exhibit the feature maps without and with MM. The model using MM pays more high-fidelity attention to details.}
	\label{fig:fea}
	\vspace{-5mm}
\end{figure}
\noindent\textbf{Mask-guided Mechanism.} We conduct ablation studies to investigate the effect of the previous mask usage scheme described in Sec.~\ref{sec:mask}, our MM, and their interaction. The adopted network is the baseline model using S-MSA. The results are reported in Tab.~\ref{tab:mask}.  Method A uses our input setting. Method B exploits the previous scheme that adopts $\mathbf{H}\odot\mathbf{M}^*$ as the input. B achieves a limited improvement due to the HSI representation corruption and under-utilization of the mask. Method C uses our MM. C yields the most significant improvement by 1.08 dB, showing the guidance advantage of MM for HSI reconstruction. D exploits both the previous scheme and our MM but degrades by 0.19 dB when compared to method C. This degradation may stem from the loss of some input spectral information.

Additionally, to intuitively show the advantages of MM, we visualize the feature maps of the last MSAB in MST-S. As depicted in Fig.~\ref{fig:fea}, the top row shows the original RGB images. The middle and bottom rows respectively exhibit the feature maps without and with MM. It can be clearly observed that the model without MM generates blurred, distorted, and incomplete feature maps while sacrificing some details, neglecting some scene patches, or introducing unpleasant artifacts. In contrast, the model using our MM pays more accurate and high-fidelity attention to the detailed contents and structured textures of the desired scenes.

\vspace{-3.5mm}
\section{Conclusion}
\vspace{-1.5mm}
In this paper, we propose an efficient Transformer-based framework, MST, for accurate HSI reconstruction. Motivated by the HSI characteristics, we develop an S-MSA to capture inter-spectra similarity and dependencies. Moreover, we customize an MM module to direct S-MSA to pay attention to spatial regions with high-fidelity HSI representations. With these novel techniques, we establish a series of extremely efficient MST models. Quantitative experiments demonstrate that our method surpasses SOTA algorithms by a large margin, even requiring significantly cheaper Params and FLOPS. Qualitative comparisons show that our MST achieves more visually pleasant HSI reconstruction results.

{\small
\bibliographystyle{ieee_fullname}
\bibliography{reference}
}

\end{document}